\documentstyle[12pt,psfig]{article}  
\begin{document}
\date{}                  
\bibliographystyle{unsrt}

\title{Equilibrium size of large ring molecules}
\author{J.M.Deutsch \\
University of California, Santa Cruz, U.S.A.}
\maketitle
%\noreportnumber
\abstract
{
The equilibrium properties of isolated ring molecules were investigated
using an off-lattice model with no excluded volume but with  dynamics that
preserve the topological class. Using an efficient set of
long range moves, chains of more than 2000 monomers were studied.
Despite the lack of any excluded volume interaction, the radius of gyration
scaled like that of a self avoiding walk, as had been previously conjectured.
However this scaling was only seen for chains greater than 500 monomers.
}

\vskip 0.3 truein    

\newpage

\section{Introduction}

The effects of topology on the equilibrium properties of polymer systems
is important both experimentally and theoretically to the understanding of
polymeric materials. Many experimental systems, for example  as polymer networks,
possess non-trivial topology. Furthermore DNA is often in the form of a
ring and possesses a well defined and non-trivial topology. This fact
has important biological implications\cite{wang,wasserman}.  Topological
effects arise because a polymer chain cannot cross itself and is
therefore confined to a fixed topological class that depends on its
initial configuration. However the effects of topology are, in theory,
very hard to understand. The simplest example is that of the ring polymer. Even
when it is in the state of a ``trivial" knot, that is the same topological class as a
circle, its statistical properties have not as yet been well understood.
This is because there are an infinite number of invariants that
characterize a knot and so evaluating, or even writing down, the partition
function is a daunting task.  Questions about this system, such as
how radius of gyration $R_g$ scales with chain length $N$, have not
been answered with any certainty.  The effects of topology on the size
of a simple ring has been conjectured to give the same scaling exponent
$R_g \sim N^\nu$, as excluded volume\cite{descloizeaux} chains.
It is not clear that $\nu$ should be identical to that of the self
avoiding chain. Topological effects may induce effective power law
interactions between monomers in the ring.
In this paper, we wish to analyze how topology alters the fractal dimension
of a polymer ring. 

To simulate polymers, lattice models have often been employed because in
many circumstances they are more efficient and easier to implement than
off-lattice models.  However a large class of lattice models with local
dynamics have been proven to be strongly non-ergodic{sokal}.  Even if an ergodic
model is used{quake} some of the subtle effects of topology that we
wish to examine here can be masked
by this kind of model.  To avoid segments crossing, which would violate topology,
lattice models add in a hard core repulsion. This makes analysis of the
results difficult because this repulsive potential must also be
taken into account.  The same can be said for many off-lattice models\cite{bishop,sheng} that
include repulsive potentials between the monomers of the ring. Again
this has the problem of masking delicate topological effects making results
difficult to interpret.

Therefore it would seem worthwhile to
explore an off-lattice model whose interactions are purely topological.
The idea is to isolate topological effects to gain some understanding
as to how they effect statistical properties.

To achieve this goal we explore topological effects numerically using an
efficient off-lattice algorithm which {\em has no excluded volume}. The
chain is a phantom chain, except that it is not allowed to cross itself
and therefore can explore only one topological class. As mentioned above, this differs
from other work in this area where excluded volume was also present. An
efficient long range algorithm is developed that allows the exploration
of rings up to $2048$ links. In addition, the chain is more flexible than
chains on a cubic lattice which as we will see is important in light of
the results that were obtained.

What was found is rather surprising. We did find that
phantom ring polymers confined to the ``trivial knot" are swollen with respect to
phantom linear chains. But even with these very flexible chains,
the effect of topology is rather weak, and rings must be longer than
$500$ links before the scaling exponent $\nu$ appears to reach that of
a self avoiding walk (SAW).

\section{The Model}

\begin{figure}[tbh]
\begin{center}
\                
\psfig{file=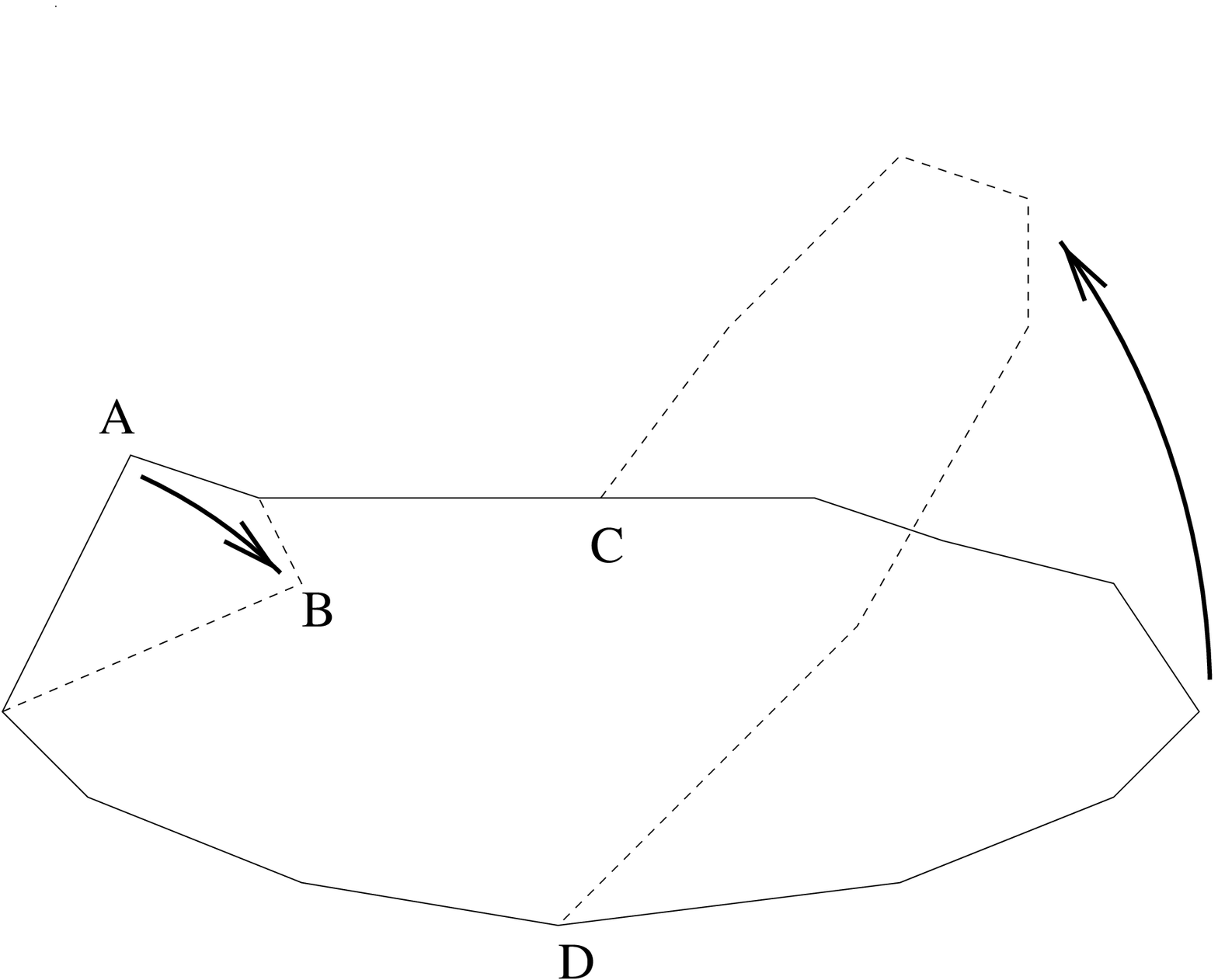,width=3in}
\end{center}
\caption{Moves illustrating the model used. A short range "kink-jump" move
is depicted in the left hand side of this figure. A single monomer is moved
from position A to position B. The right hand side illustrates a long range
move, with the dashed line being the final position of the chain.}
\label{fig:move}
\end{figure}

The model is illustrated in figure \ref{fig:move}. A chain in three dimensions
of $N$ links, all of equal length, is constructed so as to have the
topology of a simple ring. All configurations are of equal energy, however
moves must not be allowed to alter the topological class. That is, the
chain is not allowed to cross itself. 

To investigate the dynamics of small chains, moves were
chosen that were local, involving rotations of two adjacent links. Moves
of this type are illustrated in figure \ref{fig:move}, where links
initially adjacent to A are moved to locations shown by the dashed lines by B.
The relaxation time with such moves is slightly greater than $N^2$ which is
prohibitively long for chain lengths in the thousands. Therefore
a long range algorithm was used instead.

A long range move is illustrated  in the same figure. It is a variation
of a pivot algorithm\cite{madras}. To construct a new conformation, 
the algorithm attempts to rotate
the chain about some axis.  Two points C and D are chosen randomly from 
the backbone of the chain and the line going through C and D is defined to be
the axis of rotation. An angle is chosen randomly between $-90$ and
$90$ degrees and all chain segments between C and D are  rigidly rotated about 
this axis. If the ring is found to cross itself as chain segments are being
rotated, the move is rejected, otherwise it is accepted\cite{rensburg}. 

These long range moves are repeated
millions of times and the radius of gyration was averaged. 

The simulation was tested by starting two rings in a locked configuration
and checking to see that they never became unlocked. This test was performed
for over $10^8$ cycles and always remained locked. 
The simulation was further tested by taking a long
ring of length 512, after it had moved many million cycles, and turning on
a repulsive coulomb potential. It was always found that the ring would
swell up into what could clearly be seen as being a circle, 
in other words a trivial knot.
This contrasted with what happened when a deliberately knotted ring was
swollen. There the final configuration was not circular.

The simulation was checked by allowing the ring to cross itself by
turning off the checking algorithm.
In such a case it is easy to show that $R_g^2 = N/4$, when $N >> 1$. The
results found agreed with this value within statistical error. With the
crossing constraint enforced, the radii of gyration were always larger than 
this value. 

\section{Results}

The radius of gyration defined as
\begin{equation}
R_g^2 \equiv {2\over N} \sum_{i=0}^{N/2-1} |{\bf r}_{i+N/2}-{\bf r}_i|^2
\end{equation}
was examined by averaging over millions of iterations.
Here $\bf r_i$ is the vector position of the ith monomer, and N is the total
number of links in the ring.

\begin{figure}[tbh]
\begin{center}
\                
\psfig{file=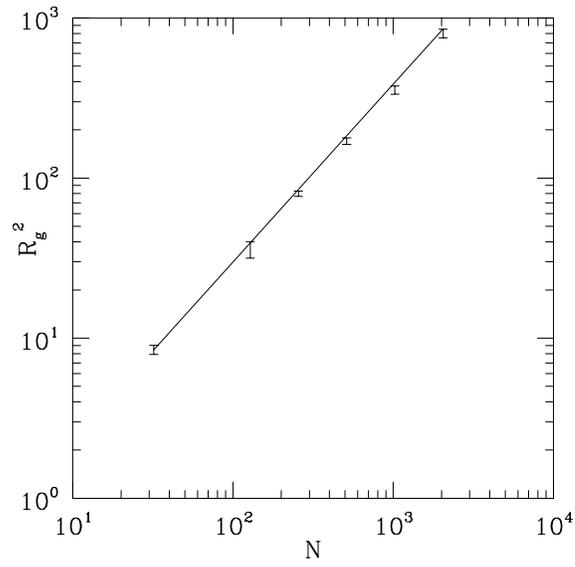,width=3in}
\end{center}
\caption{The radius of gyration as a function of chain length of rings in the class of a
trivial knot. The solid line has a slope of 1.11}
\label{fig:r_g}
\end{figure}

Figure \ref{fig:r_g} shows the $R_g^2$ as a function of chain length.
The best fit to this line gives an exponent of $1.11 \pm 0.03$.
However the best fit for the last two data points at $N ~=~ 1024$ and $N ~=~ 2048$
is an exponent of $1.17$ close to the result $1.175$ found for an
SAW in three dimensions\cite{li}.
Because this only involves two data points, the up turn is
not statistically significant so we now further analyze this 
possibility by probing the internal structure of the rings.

\begin{figure}[tbh]
\begin{center}
\                
\psfig{file=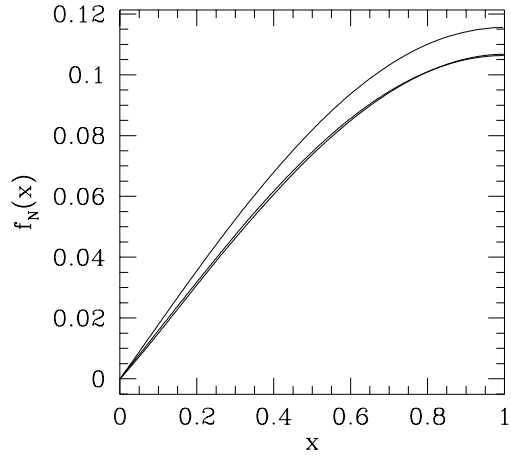,width=3in}
\end{center}
\caption{The scaling function $f_N(x)$ for three ring lengths $512$, $1024$ and $2048$. 
Here $\nu$ was chosen to be $1.17$ and the two larger ring lengths overlap.}
\label{fig:g(s)}
\end{figure}

The self similarity of these rings was examined by means of the correlation
function
\begin{equation}
g(s) \equiv {1\over N-s} \sum_{i=0}^{N-s-1} |{\bf r}_{i+s}-{\bf r}_i|^2
\end{equation}
Figure \ref{fig:g(s)} examines how this scales with chain length by writing
$g(s)$ in the scaling form
\begin{equation}
g(s) = N^\nu f_N(2s/N)
\end{equation}
The function $f_N(x)$ should become independent of $N$ for large enough $N$.
In fig \ref{fig:g(s)}, we plot $f_N(x) \equiv  g(s)/N^\nu$ as a function of
$x \equiv 2s/N$. This is done for three different chain lengths, $N=$ $512$, $1024$, and $2048$. 
$\nu$ was chosen to be $1.17$. Note that only the two longest chain lengths overlap.
In order to get the $512$ link ring to overlap with $N=2048$ , a smaller value, $\nu = 1.11$
must be chosen. This suggests that the rings of length $512$ are not long enough
to be in the asymptotic scaling regime.

\section{Discussion}

The results above confirm the prediction that $\nu$ for a polymer ring
in a trivial knot, is swollen compared to
an ideal chain. What is surprising is that the length of chain needed to
clearly see its asymptotic fractal dimension are in the thousands of links. This
contrasts with a self avoiding walk, where $\nu$ can be calculated from
chain lengths in the teens to quite high accuracy.

Fairly compelling arguments have been given to why $\nu_{ring}$ for a trivial
ring should be greater than or equal to $\nu_{SAW}$  for a linear SAW\cite{descloizeaux}.
It is still an open question as to whether $\nu_{ring}$ is actually {\em greater} than
$\nu_{SAW}$. There is no answer to this that is available
at present. However if the ring is trapped between to parallel plates,
then this is amenable to a fairly convincing analytical argument.  It is
argued here that when the separation between the plates is very small,
so that the walk is almost two-dimensional, that the exponent must be
less than or equal to $\nu$ for an SAW in two dimensions. This taken with
the argument that $\nu_{ring} \ge \nu_{SAW}$, which carries over to
this case also, we can conclude that $\nu_{ring} = \nu_{SAW}$.

If the separation between the two walls is much less than the length
of a link, then the chain is essentially confined to a two dimensional
x-y plane.  However besides the coordinates of the monomers we also must
keep track, in a crossing between two links, of which one lies on top.
If we want the ring to preserve the initial topology of a trivial knot,
some configurations where the chain crosses itself in the x-y plane
must be excluded. Essentially the only configurations that are allowed
are those of a flattened trivial knot. Therefore the phase-space of all
configurations are the coordinates in the x-y plane, and the type of
crossing of which there are two possibilities.

If in addition to the topological constraint, we introduce a fictitious
repulsive potential between links in the chain, it seems highly plausible
that this can only increase the average radius of gyration of this
system. In the limit of complete excluded volume, no crossing is allowed,
and this system is identical to the usual self avoiding ring. This is
because complete excluded volume implies that there is no crossing allowed
at all, and this by itself enforces the topological constraint. Therefore
the radius of gyration of the ring without the repulsive potential should
be less than or equal to that of a two dimensional SAW.

In three dimensions the numerical results presented here are consistent
with an exponent less than or equal to that of an SAW. However one
cannot exclude the possibility that at even longer chain lengths,
a ring becomes more swollen than a linear SAW. Until a better theoretical
understanding of knots has been achieved, it is not possible to exclude
this possibility.

It is surprising that scaling behavior has not yet reached asymptotic
behavior for chains
of length $512$ as the model we have used
allows for the formation of nontrivial knotted chains even for rings of  $6$
monomers.
However this work is consistent with enumeration of knot types for random
phantom rings\cite{sumners,koniaris} where the mean number steps needed to
observe a nontrivial knot was of order hundreds of steps.

From an experimental view point, the rings analyzed here are large
but easily achievable experimentally. A linear chain in good solvent
could be synthesized with functional end-groups that chemically bond
in good solvent. This should create a rings that are predominantly
trivial knots. By changing solvent conditions to those of theta
solvent for linear chains, one should observe that these rings are
still swollen, albeit far less so than a linear chain in good solvent.

Ring polymers in other situations\cite{catesdeutsch}, are conjectured to have many
interesting properties. For example, a melt of rings, and this is the
subject of active investigation\cite{muller}. It might be useful to apply the model
described here, to these other systems.

\section{Acknowledgements}
This work is supported by NSF grant number DMR-9419362
and acknowledgment
is made to the Donors of the Petroleum Research Fund, administered
by the American Chemical Society for partial support of this research.

\newpage                               

\newpage                               

\newpage                               

\end{document}